# Exploring the Determinants of Capital Adequacy in Commercial Banks: A Study of Bangladesh's Banking Sector

Md Shah Naoaj

## ABSTRACT

This study investigates the factors that influence the capital adequacy of commercial banks in Bangladesh using panel data from 28 banks over the period of 2013-2019. Three analytical methods, including the Fixed Effect model, Random Effect model, and Pooled Ordinary Least Square (POLS) method, are employed to analyze two versions of the capital adequacy ratio, namely the Capital Adequacy Ratio (CAR) and Tier 1 Capital Ratio. The study reveals that capital adequacy is significantly affected by several independent variables, with leverage and liquidity risk having a negative and positive relationship, respectively. Additionally, the study finds a positive correlation between real GDP and net profit and capital adequacy, while inflation has a negative correlation. For the Tier 1 Ratio, the study shows no significant relationship between leverage and liquidity risk, but a positive correlation with the number of employees, net profit, and real GDP, while a negative correlation with size and GDP deflator. Pooled OLS analysis reveals a negative correlation with leverage, size, and inflation for both CAR and Tier 1 Capital Ratio, and a positive correlation with liquidity risk, net profit, and real GDP. Based on the Hausman test, the Random Effect model is deemed more suitable for this dataset. These findings have important implications for policymakers, investors, and bank managers in Bangladesh by providing insights into the factors that impact the capital ratios of commercial banks.

**M. S. Naoaj\***
Graduate Student, Department of Economics, New York University, USA.
(e-mail: mn3116@nyu.edu)

*\*Corresponding Author*

## I. Introduction

The capital adequacy ratio (CAR) is an essential measure of a bank's financial health, indicating its ability to absorb potential losses and meet its financial obligations. The importance of maintaining an adequate CAR has been highlighted in recent years, especially with the implementation of Basel III regulations that require banks to maintain a minimum CAR of 8%. The banking sector in Bangladesh has undergone significant changes in recent years, including the introduction of new banking regulations and increased competition. In this context, it is crucial to explore the factors that determine the CAR for banks operating in Bangladesh.

While several studies have examined the determinants of CAR in other countries, there is a lack of research on this topic in the context of Bangladesh. Therefore, this paper aims to fill this research gap by analyzing the factors that influence CAR for banks operating in Bangladesh. The study will be based on data collected from the financial statements of commercial banks in Bangladesh over a specific period. Overall, this study aims to contribute to the existing literature on the determinants of CAR in the banking sector, with a specific focus on the context of Bangladesh. The findings of this study can inform policymakers and banking industry stakeholders on the key factors that impact CAR and guide them in developing policies that promote financial stability and resilience in the banking sector.

## II. Literature Review

In earlier research, it has been shown that the capital structure of banks is not solely determined by regulations (Schaek & Cihak, 2012). Juca *et al.* (2012b) found that the top 20 banks in Brazil and globally maintain capital levels at 18% and 14%, respectively, which is higher than the minimum Basel requirement of 8%. Similarities between the capital structures of banks and non-financial firms have also been observed (Gropp & Heider, 2010; Juca *et al.*, 2012a). Various factors, including profitability, growth, asset structure, size, competition, risk, and efficiency, have been identified as determinants of capital structure (Amidu, 2007; Altunbas *et al.*, 2007; Buyuksalvarci & Abdioglu, 2011; Mili *et al.*, 2014; Ogere *et al.*, 2013; Williams, 2011). The determinants of capital structure in developing economies have also been studied, with some focusing on macroeconomic factors (Octavia & Brown, 2009) and others considering environmental variables such as Eurozone stock market volatility index and governance indicators (Amjad & Tufail, 2013; Saeed *et al.*, 2013; Romdhane, 2012; Bokhari *et al.*, 2012; Mathuva, 2009; Asarkaya & Ozcan, 2007).

Aspal *et al.* (2014) conducted a study in India and found that the capital adequacy ratio (CAR) is negatively related to loan asset ratio (LAR), asset quality, and management efficiency, while liquidity and sensitivity are positively related to CAR. The study also revealed that private sector banks in India maintain a higher level of capital requirements than required by the Reserve Bank of India. Shingjergji and Hyseni (2015) found that CAR is not correlated with return on assets (ROA) and return on equity (ROE), but is negatively impacted by size, nonperforming loans (NPL), loans to deposits ratio (LTD), and equity multiplier (EM). Nonetheless, the funding structure involving capital augmentation raises the cost of equity, as estimated by Naoaj and Hosen (2023), who found that a 10 percent increase in capital would lead to a 4.39 percent reduction in the cost of equity. Ozili (2015) conducted an empirical study and discovered that bank profitability is significantly influenced by bank capital adequacy, which is a crucial determinant of bank profitability.

Thoa *et al.* (2020) investigated the determinants of Vietnamese banks' CAR using secondary data from banks' annual reports during 2009-2015. The study used panel data and found that bank size and liquidity had a significant negative impact on CAR, while loan loss reserve and loans had a negative impact but were statistically insignificant. Büyükşalvarcı and Abdioğlu (2011) investigated the determinants of CAR in Turkish banks and its impact on the financial positions of banks. The study utilized panel data analysis to explore the relationship between CAR and several independent variables, including bank size, deposits, loans, loan loss reserve, liquidity, profitability, net interest margin, and leverage, using data from banks' annual reports from 2006 to 2010. The research findings indicate that loan loss reserves and return on assets have a positive effect on CAR, while loans, return on equity, and leverage have a negative influence. However, bank size, deposits, liquidity, and net interest margin do not have a significant impact on CAR. The study highlights the significance of maintaining sufficient capital levels in Turkish banks for financial stability (Büyükşalvarcı & Abdioğlu, 2011). According to Naoaj (2019), economic crises resulting from natural disasters can influence the financial sector indicators such as capital adequacy.

In 2014, Polat and Al-Kalaf conducted a study on the determinants of capital adequacy ratio (CAR) in the banking system of the Kingdom of Saudi Arabia from 2008 to 2012. They found that all independent variables, except nonperforming loans, had a significant impact on CAR. Specifically, loans to assets ratio had a negative impact, while size and leverage had a positive impact. Loans to deposits ratio had a negative impact, while return on assets (ROA) had a positive impact on CAR. Other studies have investigated the determinants of CAR in Islamic banks. For instance, Abusharaba *et al.* (2013) studied CAR in Indonesia and found that liquidity and ROA were positively related to CAR, while nonperforming loans had a negative correlation. Abdul Karim *et al.* (2013) found a positive relationship between CAR and deposit and loan growth in both Islamic and conventional banks.

Similarly, Romdhane *et al.* (2012) reported that the determinants of CAR in developed countries are similar to those in developing countries. They analyzed the determinants of CAR in Tunisia and found that interest margin and risk had a strong impact on the capital ratio. Atici and Gursoy (2013) examined the determinants of CAR in the Turkish banking system and found that Turkish banks effectively used the capital buffering approach proposed under Basel III. Ahokpossi *et al.* (2020) suggested enhancing policy communication, such as by improving the transparency and coherence of policy messages with the policy framework to improve compliance of regulatory requirements.

Despite the impressive economic growth achieved by Bangladesh in recent years, driven by a fast-growing manufacturing sector, robust remittance inflows, and ongoing reforms, with an 8.15% GDP growth rate recorded in 2019 and nearly 7% growth over the last decade (Naoaj *et al.*, 2021), Bangladesh banking industry is facing problem with low capital adequacy with high non-performing loans. Bangladesh Bank, the central bank of Bangladesh, has taken several measures to improve capital adequacy in the banking sector. To illustrate, the central bank has introduced the minimum CAR requirement, required capital against country risk and introduced pillar 2 capital requirements (BB Basel III Policy, 2014; Country Risk Policy, 2021). Despite the extensive research on the determinants of CAR in both developed and developing economies, no previous study has examined the factors that influence capital adequacy ratios in Bangladesh. Therefore, this paper aims to fill this research gap by analyzing the factors that influence CAR for banks operating in Bangladesh.

### III. DATA DESCRIPTION

The analysis in this research is based on the data obtained from 28 commercial banks in Bangladesh, covering the period from 2013 to 2019, excluding the year 2020 due to the overshadowing effect of Covid-related uncertainty and strong fiscal and monetary measures on liquidity risk. The study gathered data from the banks' annual reports, the website of the Bangladesh Bank, and the World Bank. In this study, two definitions of capital adequacy are used, which are shown in (1) and (2).

$$CAR = \frac{\text{Core Capital (tire 1)} + \text{additional capital (tire 2)}}{\text{Risk Weighted Assets (RWA)}} \quad (1)$$

$$Tier1\_Ratio = \frac{\text{Core Capital (tire 1)}}{\text{Risk Weighted Assets (RWA)}} \quad (2)$$

Table I presents summary statistics for 13 variables related to the banking sector. The definition of the variables with their economic meaning is given below:

- leverage: This is the ratio of its total loans to its total assets. A higher leverage ratio means that the bank is financing a larger portion of its assets with debt, which can increase profitability in good times but also increase the risk of insolvency in bad times.
- liq_risk: Liquidity risk is total loans divided by total deposits, which is a measure of its ability to meet funding needs. Higher liquidity risk means that the bank is more likely to experience funding shortfalls.
- l_employee: This is the natural logarithm of a bank's number of employees. A larger number of employees may indicate higher operating costs and lower efficiency, but it may also reflect greater specialization and more diverse business lines.
- l_bad_loan: This is the natural logarithm of a bank's

total non-performing loans. Higher levels of non-performing loans can indicate greater credit risk and lower profitability, as the bank may need to set aside reserves to cover potential losses.
- l_size: This is the natural logarithm of a bank's total assets. Larger banks may benefit from economies of scale, but they may also face greater regulatory scrutiny and systemic risk concerns.
- l_net_profit: This is the natural logarithm of a bank's net profit, which is its total revenue minus all expenses. Higher net profits may indicate better financial health and greater efficiency.
- l_staff_exp: This is the natural logarithm of a bank's staff expenses, which include salaries, bonuses, and other compensation for employees. Higher staff expenses may indicate a larger workforce or more generous compensation, which can impact the bank's profitability.
- l_op_exp: This is the natural logarithm of a bank's operating expenses, which include all non-interest expenses such as salaries, rent, and utilities. Higher operating expenses may indicate higher levels of inefficiency and lower profitability.
- real_gdp: This is the real gross domestic product, which is a measure of the total economic output adjusted for inflation. A higher real GDP may indicate greater economic activity and more favorable business conditions for banks.
- gdp_deflator: This is the GDP deflator, which is a measure of the general price level of all goods and services in an economy. A higher GDP deflator may indicate inflationary pressures, which can impact the profitability and risk of banks.

The first column ("Obs") indicates the number of observations for each variable. The second column ("Mean") shows the average value of each variable across all observations. The third column ("Std. Dev.") shows the standard deviation of each variable, which is a measure of how much the values vary from the mean. The fourth column ("Min") shows the minimum value for each variable, while the fifth column ("Max") shows the maximum value.

TABLE I: SUMMARY STATISTICS

| Variable | Obs | Mean | Std. Dev. | Min | Max |
|---|---|---|---|---|---|
| car | 196 | 0.127 | 0.015 | 0.1 | 0.18 |
| tier1 ratio | 196 | 0.092 | 0.015 | 0.06 | 0.14 |
| leverage | 196 | 0.662 | 0.065 | 0.472 | 0.837 |
| liq risk | 196 | 0.881 | 0.089 | 0.571 | 1.066 |
| l employee | 196 | 7.965 | 0.533 | 7.214 | 9.64 |
| l bad loan | 196 | 6.601 | 0.555 | 5.154 | 8.143 |
| prov | 195 | 0.209 | 0.114 | -0.01 | 0.63 |
| l size | 196 | 10.071 | 0.428 | 9.098 | 11.645 |
| l net profit | 196 | 5.326 | 0.651 | 1.085 | 6.527 |
| l staff exp | 196 | 5.549 | 0.544 | 3.267 | 7.285 |
| l op exp | 196 | 6.96 | 0.429 | 5.924 | 8.418 |
| real gdp | 196 | 223.673 | 51.217 | 150 | 302.57 |
| gdp deflator | 196 | 196.42 | 22.342 | 164.26 | 229.41 |

Table II is a correlation matrix that shows the correlation coefficients between each pair of variables. The correlation coefficient measures the strength and direction of the linear relationship between two variables, ranging from -1 to 1. A value of 1 indicates a perfect positive correlation, while a value of -1 indicates a perfect negative correlation, and a value of 0 indicates no correlation.

The table also shows the strength of the correlations, which can be interpreted as follows: a coefficient between 0.1 and 0.3 indicates a moderate correlation, while a coefficient above 0.3 indicates a strong correlation. It is important to note that a correlation does not imply causation, and other factors may be influencing the relationships between the variables.

TABLE II: CORRELATION MATRIX

| Variables | (1) | (2) | (3) | (4) | (5) | (6) | (7) | (8) |
|---|---|---|---|---|---|---|---|---|
| (1) car | 1.0 | - | - | - | - | - | - | - |
| (2) tier1_ratio | 0.5 | 1.0 | - | - | - | - | - | - |
| (3) leverage | -0.2 | -0.3 | 1.0 | - | - | - | - | - |
| (4) liq_risk | 0.1 | -0.2 | 0.8 | 1.0 | - | - | - | - |
| (5) l_employee | 0.2 | 0.2 | 0.0 | -0.1 | 1.0 | - | - | - |
| (6) l_bad_loan | 0.2 | 0.0 | 0.3 | 0.3 | 0.6 | 1.0 | - | - |
| (7) l_size | 0.2 | -0.1 | 0.3 | 0.3 | 0.7 | 0.8 | 1.0 | - |
| (8) l_net_profit | 0.4 | 0.3 | 0.0 | 0.1 | 0.5 | 0.4 | 0.5 | 1.0 |
| (9) l_staff_exp | 0.3 | 0.1 | 0.0 | 0.1 | 0.7 | 0.6 | 0.7 | 0.5 |

## IV. METHODOLOGY

The methodology used in this study involved the analysis of panel data for the banking sector of Bangladesh. Panel data analysis is a statistical method used to analyze data that contains both cross-sectional and time series components. In this study, the panel data consisted of data from multiple banks over a period of time.

Previous research has identified various methods for analyzing panel data, which include Pooled Ordinary Least Square (POLS), fixed-effect model, and random effect model, among others (Greene, 2012). Typically, when the number of independent variables is fixed, and all variables are expressed in ratios, the fixed-effect model is preferred for regression and variance analysis (Wooldridge, 2002). However, in the present study, while the number of independent variables is fixed, not all terms are represented in ratios. Therefore, to determine whether the Random Effect Model is more suitable for this dataset, the Hausman test is conducted (Hausman, 1978). Furthermore, to ensure the robustness of the findings, the study also employed the Pooled Ordinary Least Square (POLS) method.

Overall, the combination of panel data analysis and descriptive statistics provided a robust methodology for examining the determinants of the capital adequacy ratio for the banking sector of Bangladesh.

The regression models are represented as follows:

$$car = \alpha + \beta1 leverage_{it} + \beta2 liq\_risk_{it} + \beta3 l\_bad\_loan_{it} + \beta4 l\_bank\_size_{it} + \beta5 l\_net\_profit_{it} + \beta6 l\_staff\_exp_t + \beta7 l\_op\_expense_t + U_i + e_{it} \quad (3)$$

$$tier1\_ratio = \alpha + \beta1 leverage_{it} + \beta2 liq\_risk_{it} + \beta3 l\_bad\_loan_{it} + \beta4 l\_bank\_size_{it} + \beta5 l\_net\_profit_{it} + \beta6 l\_staff\_exp_t + \beta7 l\_op\_expense_t + U_i + e_{it} \quad (4)$$

The equations are panel data regression model, where the dependent variable is capital adequacy ratio (car) or the tier 1 capital ratio, and the independent variables are leverage, liquidity risk, bad loans, bank size, net profit, staff expenses, and operating expenses. The model also includes fixed effects (Ui) and a random error term (e_it). The parameters of the model are represented by β1 to β7 and α is the intercept. The independent variables (leverage, liquidity risk, bad loans, bank size, net profit, staff expenses, and operating expenses) are measured at time t for each bank i. The fixed effects (Ui) are the unobserved time-invariant characteristics that affect the tier 1 capital ratio of each bank i. The fixed effects capture the unobserved heterogeneity across the individual banks that do not change over time. The random error term (e_it) captures the random shocks that affect the tier 1 capital ratio of each bank i at each time t, which is assumed to be independent and identically distributed across time and banks.

## V. RESULTS

Table III shows the estimated coefficients for several independent variables and two dependent variables - car and tier1 capital ratio - using three different regression methods - Pooled OLS, Fixed Effects, and Random Effects. The Hausman test confirms that the random effect model is consistent. The coefficients represent the direction and strength of the relationship between each independent variable and the dependent variable, with a positive sign indicating a positive relationship, and a negative sign indicating a negative relationship. The size of the coefficient indicates the strength of the relationship, with larger values indicating a stronger relationship. The p-values associated with each coefficient indicate whether the relationship between the independent and dependent variables is statistically significant or not. If the p-value is less than 0.1, 0.05, or 0.01, then the relationship is considered statistically significant at the 10%, 5%, or 1% level, respectively.

Looking at the table, we can see that some variables have statistically significant relationships with the dependent variables, while others do not. For example, in the Pooled OLS model, the leverage variable has a negative and statistically significant relationship with both car and tier1 capital ratio, while the l_size variable has a negative and statistically significant relationship with car but not with tier1 capital ratio. As Hausman test suggests, the random effect model is preferred over fixed effect model, lets explain the outcome of the random effect model. The random effect results show that leverage has a negative and statistically significant relationship with car, indicating that as leverage increases, capital adequacy decreases. Liquidity risk has a positive and statistically significant relationship with car and tier1 capital ratio, suggesting that as liquidity risk increases, capital adequacy also increases.

Log of total employees (l_employee) has a positive and statistically significant relationship with tier1 capital ratio (coefficient = 0.0118, p<0.01), but not with car. Log of net profit (l_net_profit) has a positive and statistically significant relationship with car (coefficient = 0.00548, p<0.01) and tier1 capital ratio (coefficient = 0.00916, p<0.05), indicating that as net profit increases, capital adequacy also increases.

Log of total assets (l_size) has a negative but statistically insignificant relationship with and tier1 capital ratio, suggesting that there is no significant relationship between the two variables. Log of non-performing loans (l_bad_loan), log of staff expenses (l_staff_exp), and log of operating expenses (l_op_exp) have no statistically significant relationship with car or tier1 capital ratio.

Real gross domestic product (real_gdp) has a positive and statistically significant relationship with both car and tier1 capital ratio, indicating that as real GDP increases, capital adequacy also increases. Gross domestic product deflator (gdp_deflator) has a negative and statistically significant relationship with both car and tier1 capital ratio, suggesting that as inflation increases, capital adequacy decreases.

TABLE III: RESULTS

| VARIABLES | Pooled OLS | | Fixed Effect | | Random Effect | |
|---|---|---|---|---|---|---|
| | (1) car | (2) tier1 ratio | (3) car | (4) tier1 ratio | (5) car | (6) tier1 ratio |
| leverage | -0.103*** | -0.0653** | -0.155*** | -0.00335 | -0.118*** | -0.0409 |
| | (0.0281) | (0.0255) | (0.0513) | (0.0454) | (0.0373) | (0.0324) |
| liq_risk | 0.0647*** | 0.0432** | 0.0673* | -0.0170 | 0.0662** | 0.0112 |
| | (0.0185) | (0.0184) | (0.0340) | (0.0316) | (0.0288) | (0.0258) |
| l_employee | 0.00263 | 0.0120*** | 0.0107 | -0.00170 | 0.00581 | 0.0118*** |
| | (0.00307) | (0.00317) | (0.00948) | (0.00727) | (0.00423) | (0.00401) |
| l_bad_loan | -0.000992 | 0.0100*** | -0.00470 | 0.000695 | -0.00308 | 0.00265 |
| | (0.00288) | (0.00351) | (0.00364) | (0.00444) | (0.00319) | (0.00454) |
| l_size | -0.0121* | -0.0288*** | -0.0255 | -0.0251 | -0.0102 | -0.0203* |
| | (0.00702) | (0.00770) | (0.0190) | (0.0178) | (0.00937) | (0.0105) |
| l_net_profit | 0.00724*** | 0.0115*** | 0.00452** | 0.00933** | 0.00548*** | 0.00916** |
| | (0.00151) | (0.00285) | (0.00185) | (0.00357) | (0.000995) | (0.00356) |
| l_staff_exp | 0.00275 | 0.00460* | 0.000616 | 0.00105 | 0.00142 | 0.00270 |
| | (0.00271) | (0.00247) | (0.00329) | (0.00258) | (0.00262) | (0.00297) |
| l_op_exp | 0.00287 | -0.00863 | 0.00442 | -0.0173 | 0.00124 | -0.0110 |
| | (0.00732) | (0.00797) | (0.0130) | (0.0132) | (0.00749) | (0.0103) |
| real_gdp | 0.00152*** | 0.00158*** | 0.00135*** | 0.00129*** | 0.00149*** | 0.00141*** |
| | (0.000532) | (0.000478) | (0.000366) | (0.000344) | (0.000392) | (0.000342) |
| gdp_deflator | -0.00323*** | -0.00370*** | -0.00258*** | -0.00269*** | -0.00310*** | -0.00318*** |
| | (0.00122) | (0.00110) | (0.000834) | (0.000815) | (0.000909) | (0.000796) |
| Constant | 0.466*** | 0.572*** | 0.520*** | 0.675*** | 0.455*** | 0.526*** |
| | (0.122) | (0.109) | (0.155) | (0.126) | (0.0990) | (0.0817) |
| Observations | 196 | 196 | 196 | 196 | 196 | 196 |
| R-squared | 0.406 | 0.443 | 0.362 | 0.307 | - | - |
| Number of bank | - | - | 28 | 28 | 28 | 28 |

Standard errors in parentheses. *** p<0.01, ** p<0.05, * p<0.1

Overall, the results show that several independent variables have a statistically significant relationship with car and/or tier1 capital ratio. Leverage and liquidity risk are important factors affecting capital adequacy, with a negative relationship for leverage and a positive relationship for liquidity risk. Additionally, real GDP and net profit have a positive relationship with capital adequacy, while inflation has a negative relationship with it. However, for tier 1 ratio, leverage, and liquidity risk don't show any significant relationship but a positive relationship with the number of employees, net profit, and real GDP while negative relation with size and GDP deflator. In pooled OLS, leverage, size, and inflation show negative relations for both car and tier 1 capital ratio while liquidity risk, net profit, and real GDP exhibit positive relations.

## VI. Conclusion

Based on the analysis of the panel data, the study has found that the capital adequacy ratio (CAR) of banks is positively associated with the liquidity risk and real GDP growth rate, and negatively associated with the leverage, size of the bank, and GDP deflator. Meanwhile, the tier 1 capital ratio of banks is positively related to the number of employees, net profit, real GDP growth rate, and GDP deflator, and negatively related to the size of the bank. Overall, the findings of the study suggest that the capital adequacy ratio and tier 1 capital ratio of banks are affected by various factors, and this can have implications for financial stability and risk management in the banking sector. The study recommends that policymakers should consider the factors affecting capital adequacy and tier 1 capital ratios of banks when formulating and implementing regulations and policies to promote financial stability and soundness of the banking system.

## Conflict of Interest

Author declares that they do not have any conflict of interest.

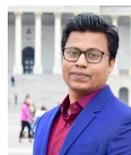

**Md Shah Naoaj** is a graduate student in NYU's Department of Economics and a Graduate Research Assistant at NYU Stern's DHL Initiative on Globalization. His research interests include international economics, financial crises, financial sector development, income and inequality, and monetary policy.

Before studying at NYU, Mr. Naoaj served the Central Bank of Bangladesh as a Deputy Director and International Monetary Fund as an Economist. He graduated from the University of Dhaka with MBA and BBA in Finance. He is also a CFA Charterholder. While at the Central Bank, he was part of a team overseeing Bangladesh's financial sector development issues and conducting relevant research. At the IMF, he was a Brunei Article IV mission team member and co-authored several papers and briefs on central bank communication, natural disasters, and financial deepening.